# *Kinetics-Optimized Enhanced Sampling Using Mean First Passage Times*

Tiejun Wei[1], Balint Dudas[1], and Edina Rosta[1]

[1]*University College London, Department of Physics and Astronomy*
*London WC1E 6BS*

*Correspondence: e.rosta@ucl.ac.uk*

Molecular dynamics (MD) simulations have become essential in many areas of atomistic modelling from drug discovery to materials science. They provide critical atomic-level insights into key dynamical events experiments cannot easily capture. Unfortunately, their impact often falls short as the timescales of the important processes are inaccessible using standard MD. Enhanced sampling methods provided avenues to access such crucial rare events, for example key slow conformational changes of biomolecules. However, the bias in enhanced sampling simulations is rarely optimized, and even if it is, the optimization criterion is based on the thermodynamics or Hamiltonian of the system, but not directly considering molecular kinetics. Here, we introduce a novel enhanced sampling algorithm that adaptively optimizes the bias based on the kinetics of the system for the first time. We identify the optimal bias that minimizes a key physical observable, the mean first passage time from a starting state to a target state. Our algorithm makes use of the relation between biased and unbiased kinetics obtained from discretized Markov state models, as established in the dynamic histogram analysis method. We demonstrate the applicability of our method for analytical potential-based model examples, NaCl dissociation in explicit water, and phosphate unbinding in Ras GTPase. Our algorithm has excellent performance compared with state-of-the-art enhanced sampling methods in terms of the timescales required to reach the final state. Our findings provide a novel, kinetics-driven enhanced sampling strategy, that uses a target-directed approach to facilitate mapping rare events, with the potential for breakthrough applications in drug discovery and materials science.

## I. INTRODUCTION

Biomolecular simulations, including all atom molecular dynamics (MD) have become essential to provide unprecedented details into the dynamics of complex systems. In particular, slow conformational changes in biomolecular systems play a crucial role in many biological functions, but are mostly unavailable experimentally, particularly the atomic level insights into the rate limiting transition states and corresponding mechanistic pathways. MD simulations therefore find key applications in e.g., drug discovery, as powerful tools to provide important structural and energetic information for complex systems of biomedical interest [1,2]. On the other hand, their impact often falls short due to the well-known sampling problem [3,4]. While aiming to uncover pivotal rare events, these simulations frequently stall in local minima, consuming extensive computational resources, limiting the timescales that can be reached to study biomolecular processes.

Enhanced sampling algorithms are developed to overcome this dilemma, and access orders of magnitude longer timescales that would be beyond reach for current computational resources [5]. A wide variety of simulation strategies were proposed to map pathways from an initial state based on adding a bias to the potential energy of the system. Beyond the classical umbrella sampling (US) [6], adaptive methods such as the finite temperature string method [7], its variant with coupled Hamiltonian replica exchange method [8], targeted MD [9], milestoning-based sampling [10], Adaptive Biasing Force [11–13], as well as metadynamics (MetaD) and its improved variants, well-tempered metadynamics (WT-MetaD) [14,15] were successful to explore the conformational space by deploying and optimizing bias [16].

In most biasing strategies, thermodynamics-based bias is applied to the system. For example, MetaD aims to explore the selected collective variable (CV) space and iteratively identify the bias that, when added to the original free energy surface, provides a flat surface [14]. WT-MetaD includes additional parameters to enable a smooth biasing process [15]. Additionally, machine learning-aided enhanced sampling methods are also gaining attention as a promising avenue for CV selection and sampling [17–21]. However, these algorithms are not directly optimal to reduce the timescale for the conformational change of interest, and the bias is not optimized for the underlying molecular kinetics.

To describe the underlying molecular kinetics of the system, Markov state models (MSM) were highly successful [3,22,23]. These also find applications in both deep learning and reinforcement learning strategies, such as using the multi-armed bandit framework [24] or autoencoders [25]. Alternatively, transition path sampling (TPS) methods [26–28] and finite temperature string method [7] also provide valuable insights into the kinetics of rare events [10,28,29]. However, the lack of bias in these methods limits the timescales that can be reached [30].

To design optimal kinetics-based bias, an analytical relationship is needed between the dynamics of the biased and unbiased systems. However, the weighted histogram analysis method (WHAM) [31], which is typically used to estimate the unbiased free energy profiles from biased runs, does not provide kinetic information. To enable the estimation of the unbiased kinetics, we use the dynamic histogram analysis method (DHAM) [22], which provides the necessary theoretical framework to build the unbiased MSM from multiple biased trajectories. The DHAM and its extension to rate matrices (DHAMed) [32] were developed to yield unbiased rates/Markov matrices, as well as the free energy surface (FES).

Here, we propose a novel enhanced sampling algorithm that identifies the optimal bias to reach the target state from an initial starting state. We use the mean first passage time (MFPT) as a key quantity that measures the time required for the transition pathway as our optimization criterion [33–35]. We formulate an optimization problem, where the bias is defined as the summation of a set of Gaussian functions (see SI.1), with the underlying parameters optimized to minimize the MFPT from the starting state to the target state. We demonstrate that such optimal biased potential leads to an overall potential energy corresponding to downhill trajectories in analytical examples. We also provide examples of all atom MD simulations of a solvated NaCl ion pair and Ras-Ras GTPase-activating protein (Ras.GAP) complex system [36] to benchmark the performance of the proposed algorithm with WT-MetaD and classical MD.

## II. MARKOV STATE MODELS AND MEAN FIRST PASSAGE TIMES

The MSM model serves as a quantitative tool to analyze the dynamics of the system as a Markov chain. It is common to use discrete states to represent different conformational clusters of the system. The transition between the states can be explicitly expressed by a rate matrix $\boldsymbol{K}$, or a transition probability matrix, $\boldsymbol{M}$, the Markov matrix. The rate matrix $\boldsymbol{K}$ contains the transition probabilities per unit time. In the context of continuous time, it describes the progress of an initial population at time 0, $\boldsymbol{p}(0) \in [0,1]^n$ to the population after a lag time $t$:

$$\boldsymbol{p}(t) = e^{\boldsymbol{K}t}\boldsymbol{p}(0) \tag{1}$$

where $\boldsymbol{p}(t)$ represents the population of $n$ discrete states at time $t$. The corresponding transition probability matrix or Markov matrix for lag time $t$ is defined as $\boldsymbol{M} = e^{\boldsymbol{K}t}$.
A key kinetic quantity of the system is the MFPT matrix, which provides a measure of how quickly the system reaches the target state $j$ starting from state $i$. [35,37]. The MFPT matrix can be calculated from $\boldsymbol{K}$ (or analogously from $\boldsymbol{M}$) as:

$$t_{ji} = \frac{1}{\boldsymbol{p}_j^{\text{eq}}}\left[(\boldsymbol{p}^{\text{eq}}\boldsymbol{1}_n^T - \boldsymbol{K})_{jj}^{-1} - (\boldsymbol{p}^{\text{eq}}\boldsymbol{1}_n^T - \boldsymbol{K})_{ji}^{-1}\right] \tag{2}$$

where $t_{ji}$ represents the MFPT from state $i$ to state $j$, $\boldsymbol{p}_j^{\text{eq}}$ stands for the equilibrium probability of state $j \in \{1,\dots,n\}$ with shape of $[n, 1]$, and $\boldsymbol{1}_n^T$ is an all-ones row vector. Eq. (2) quantifies the average time it takes for the system to first reach state $j$, starting from state $i$ [37]. Time and rates in Eq. (2) are expressed in a fixed model time unit and its inverse. For transition matrices, lag-step MFPTs are multiplied by the sampling lag.

***KINETICS-OPTIMIZED ENHANCED SAMPLING USING MFPTS***

The aim is to identify the best $u^*$ bias that would provide an optimal biased rate matrix, $\boldsymbol{K}_{bias}^{\alpha *}$, which minimizes $t_{fs}$ MFPT given a starting state, $s$, to the final state, $f$, note that $s, f \in \{1, \dots, n\}$. We envision that the optimal bias is determined iteratively based on the current sampling history, to allow progressive exploration of the phase space over several iterations, $\alpha$. We introduced a relation between unbiased and biased rate or Markov matrices [22], such that the $ji$-th element of a biased rate matrix $\boldsymbol{K}_{bias}^{\alpha}$ and unbiased rate matrix $\boldsymbol{K}$ have the following relationship:

$$\frac{k_{ji}^{\alpha}}{k_{ji}} \propto \exp\left(-\frac{u_j^{\alpha}-u_i^{\alpha}}{2k_B T}\right) \quad (3)$$

where $u_i^{\alpha}$ is the bias for state $i$ in iteration $\alpha$, $k_B$ is the Boltzmann constant, and $T$ is the temperature of the simulation. For the Markov matrix ($\boldsymbol{M}_{bias}^{\alpha}$) the analogous relation holds for short lag time with good approximation [38].

We consider the system with potential energy $U(\boldsymbol{x})$, where $\boldsymbol{x} \in \mathbb{R}^{3N}$ represents the 3D coordinates of the $N$ atoms of the system, and $U$ represents the overall potential energy. We define a Markov transition probability matrix typically in a reduced space, via a $d$-dimensional CV space: $\xi(\boldsymbol{x}) \in \mathbb{R}^d$.

As bias, in this work we use a sum of $N_{gauss}$ Gaussian functions in the CV space $\xi(\boldsymbol{x})$, with the $i$-th Gaussian kernel $G_i^{\alpha}$ centered at $\mu_i^{\alpha}$, with amplitude $A_i^{\alpha}$ and width $\sigma_i^{\alpha}$, such that:

$$u_{bias}^{\alpha}(\xi) = \sum_{i=1}^{N_{gauss}} G_i^{\alpha}(\mu_i^{\alpha}, A_i^{\alpha}, \sigma_i^{\alpha}; \xi) \quad (4)$$

The Gaussian parameters $(\mu_i^{\alpha}, A_i^{\alpha}, \sigma_i^{\alpha})$ can then be optimized such that they yield the minimal MFPT value $t_{fs}$ from the starting state $s$, to the final state $f$. The search for the optimal bias $u_{bias}^{\alpha}$ of the system is a simple minimization problem as defined below:

$$u_{bias}^{\alpha} = \underset{U_{bias}(\xi)}{\operatorname{argmin}}\left\{\text{MFPT}\left(\boldsymbol{x}_s \to \boldsymbol{x}_f\right)\middle|\boldsymbol{K}\right\} \quad (5)$$

where we optimize all $\mu_i^{\alpha}, A_i^{\alpha}, \sigma_i^{\alpha}$, $i = 1, \dots, N_{gauss}$ parameters for a given unbiased rate matrix $\boldsymbol{K}$ to minimize the mean first passage time $t_{fs}$, where $f$ and $s$ are the bins that correspond to positions $\xi(\boldsymbol{x}_f)$ and $\xi(\boldsymbol{x}_s)$, respectively.

In most applications, the FES of the system is *a priori* unknown, therefore the kinetic information needed to construct a Markov matrix has to be collected from simulation data. Starting from unbiased or weakly biased initial sampling (SI.1 and SI.5), we can iteratively update the Markov matrix of the system as well as the optimal bias. In the zero-bias formulation, we start with $u_{bias}^{\alpha}(\xi(\boldsymbol{x})) = 0$ at iteration $\alpha = 0$, yielding a new trajectory $\boldsymbol{x}^{\alpha}(t), t = 0, \dots, T_{max}$ of length $T_{max}$, where the coordinates $\boldsymbol{x}$ of the system are recorded and the corresponding CV, $\xi(\boldsymbol{x})$ is calculated. We iteratively update the estimated Markov matrix using the DHAM algorithm given all the $\alpha$ trajectories, $\xi^{i}(t), i = 0, \dots, \alpha$, and their corresponding bias [22]. This is achieved by unbiasing all the trajectories $\boldsymbol{x}^{i}(t)$ with sets of biasing potentials $u_{bias}^{i}(\xi)$ used throughout all iterations $i = 0, \dots, \alpha$, and obtaining an estimated Markov matrix $\boldsymbol{M}_{\text{est}}^{\alpha}$.

However, during most propagation steps, prior to reaching the *global target state* $\boldsymbol{x}_f$, the estimated Markov matrix $\boldsymbol{M}_{\text{est}}^{\alpha}$ does not hold the kinetic information linking any state to this global target. To address this, we introduced intermediate target states, termed a *local target state* $\boldsymbol{x}'_f$, to guide the optimization of biasing potentials. The local target is defined as the state most proximate to the *global target state* within the explored CV space. Note that if we have recorded the transition linking any state to the *global target state*, we will have $\boldsymbol{x}_f = \boldsymbol{x}'_f$ .

We then find the new set of Gaussian kernel parameters $\left(\mu_i^{\alpha+1}, A_i^{\alpha+1}, \sigma_i^{\alpha+1}\right), i = 1, \dots, N_{gauss}$ that constitute the bias $G_i^{\alpha+1}$ for the next iteration $(\alpha + 1)$ with the bias potential $u_{bias}^{\alpha+1}(\xi)$. Note that we search this by minimizing the biased MFPT from the last position, $\boldsymbol{x}^{\alpha}(T_{max})$ to the *local target state* $\boldsymbol{x}'_f$:

$$u_{bias}^{\alpha+1} = \underset{U_{bias}(\xi)}{\operatorname{argmin}}\left\{\text{MFPT}\left(\boldsymbol{x}^{\alpha}(T_{max}) \to \boldsymbol{x}'_f\right)\middle|\boldsymbol{M}_{est}^{\alpha}\right\}. \quad (6)$$

The proposed novel enhanced sampling algorithm, termed as KOMBI (Kinetics-Optimized Markovian Bias Integration), can be summarized as the pseudo code in Algorithm 1.

**Algorithm 1.** KOMBI CV Space Exploration

0: **procedure** KOMBI enhanced sampling

1: Initialize the simulation system with potential energy $U(\boldsymbol{x})$ and a starting state $\boldsymbol{x}^{\alpha}(0)$ for $\alpha = 0$.
2: Define global target state $\boldsymbol{x_f}$ and convergence threshold $\delta$.
3: Initialize Gaussian parameters for the initial biasing potential. $u^{\alpha}_{bias}(\xi) = 0$ for $\alpha = 0$.
4: **for** $\alpha = 1$ to Maxiter **do:**
5: Perform MD with potential energy $U(\boldsymbol{x}) + u^{\alpha}_{bias}(\xi)$.
6: Obtain trajectory $\boldsymbol{x}^{\alpha}(t)$ of length $T_{max}$.
7: **if** $\left(\left|\boldsymbol{x} - \boldsymbol{x_f}\right| < \delta\right.$, for any $\boldsymbol{x}$ in $\left.\boldsymbol{x}^{\alpha}\right)$:
8: Target reached
9: **break**
10: **end if**
11: Reconstruct the unbiased Markov matrix:

$$\boldsymbol{M}^{\alpha}_{est} = \text{DHAM}\left(\boldsymbol{x}^{i}(t), u^{i}_{bias}(\xi) \middle| i = 0, \ldots, \alpha\right)$$

12: Find optimal **bias** $u^{\alpha+1}_{bias}(\xi)$ to minimize MFPT to target state:

$$u^{\alpha+1}_{bias} = \underset{U_{bias}(\xi)}{\text{argmin}}\left\{\text{MFPT}\left(\boldsymbol{x}^{\alpha}(T_{max}) \to \boldsymbol{x}'_{\boldsymbol{f}}\right) \middle| \boldsymbol{M}^{\alpha}_{\text{est}}\right\}.$$

13: **end for**
14: **end procedure**

## III. RESULTS

We first demonstrate the algorithm for a 1D model potential-derived rate matrix $\boldsymbol{K}$. The original FES is constructed using an analytical potential $U$ (Fig. **1**, black line) that is discretized into $n$ states. The transition rates in the unbiased rate matrix are restricted to neighboring states only, where all the non-neighboring transition rate are set to 0:

$$K_{i\pm1,i} = A\exp\left(\frac{-\left(U(i\pm1) - U(i)\right)}{2k_BT}\right). \quad (7)$$

For the analytical illustration in Fig. 1, we minimized the exact MFPT of the nearest-neighbour rate model for four starting states and a fixed target. Each diagonal rate is minus the sum of its off-diagonal column elements. Thirty Gaussians of fixed height 2 kcal/mol were used, with centers and widths optimized as specified in SI.2. The lowest objective found in eight independent optimizations is shown for each start. The resulting bias lowers the barriers between the starting state and the target. This illustrates the MFPT objective: $\text{MFPT}\left(\boldsymbol{x}^{\alpha}(T_{max}) \to \boldsymbol{x}_f\right)$. The adaptive MD benchmarks below instead estimate the transition model from sampled trajectories; a 2D example is shown in Fig. 2.

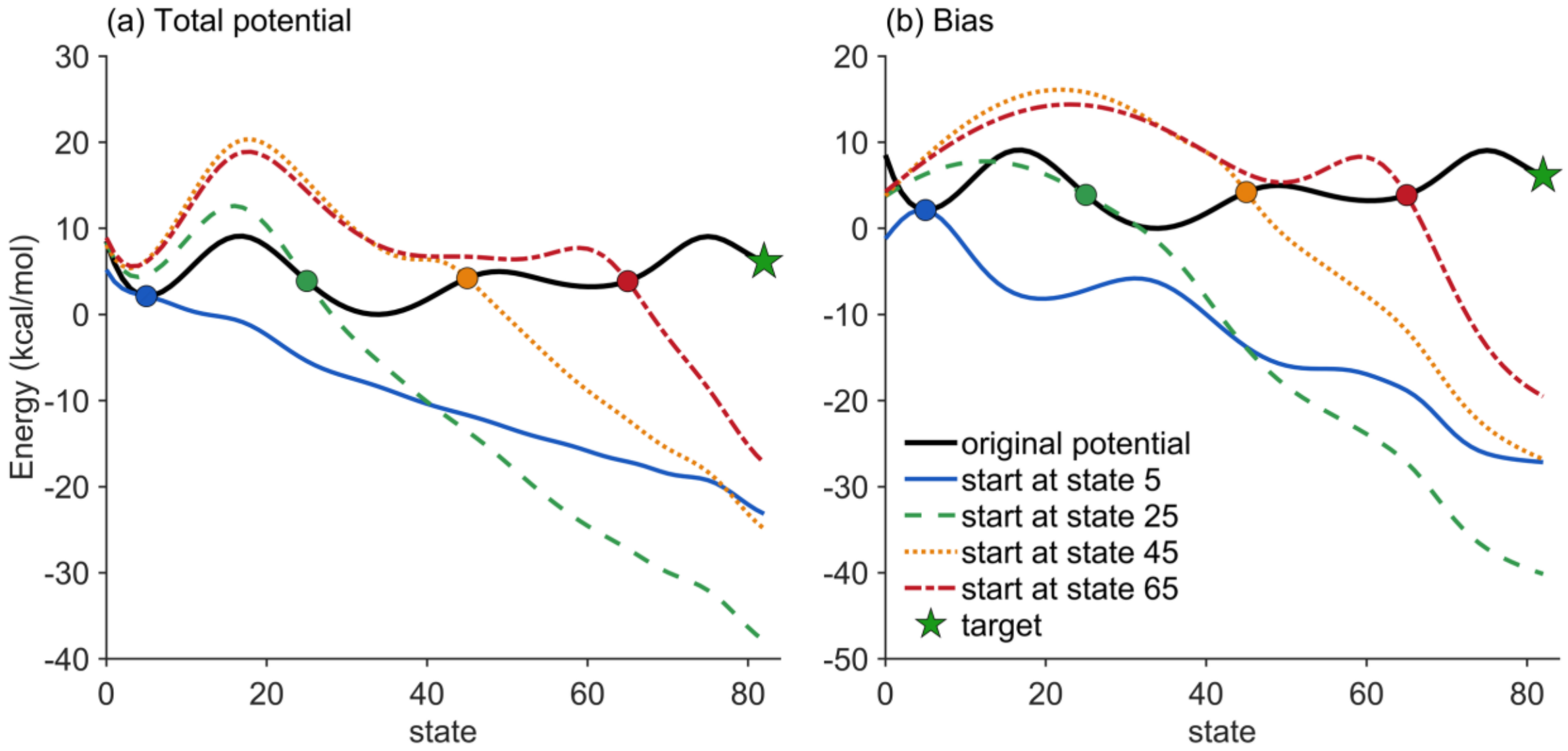

*Figure 1. Exact-potential MFPT optimization on the 1D nearest-neighbour rate model (SI.2). (a) The original potential (black) and total biased potentials for starting states 5, 25, 45 and 65 (colored circles), with target state 82 (green star). Each bias is shifted so that its value is zero at the starting state. (b) The corresponding bias functions, shifted for display to pass through the original potential at the starting state.*

We then applied Algorithm 1, assuming that the FES is not known *a priori,* in several benchmarking systems: the 1D model potential (Fig. **1**), a 2D model potential (Fig. **2**), as well as all atom simulations of solvated NaCl in a box and the dissociation of the cleaved phosphate in Ras.GAP after the GTP hydrolysis step. The simulation details for the benchmarking systems, together with additional discussion of computational cost and current limitations, can be found in SI.1–6.

For the 1D model potential, we performed 30 simulations with each method. KOMBI reached the target in 22.5 ± 0.6 ps (median 21.3 ps), compared with 1.27 ± 0.07 ns for WT-MetaD and 9.0 ± 1.1 μs for unbiased Langevin dynamics (Fig. 3). These correspond to approximately 56-fold and $4.0 \times 10^5$-fold reductions in mean target-arrival times, respectively. All times include the initial sampling period, and all 30 trajectories in each set reached the prescribed target. A representative unbiased trajectory is shown in Fig. S1.

For the 2D potential, KOMBI reached the target in 193 ± 40 ps (median 100 ps), compared with 2.45 ± 0.16 ns for WT-MetaD and 7.9 ± 1.3 μs for unbiased dynamics (Fig. 3). The mean arrival time is reduced by factors of approximately 13 and $4.1 \times 10^4$, respectively. Fig. 2 shows representative trajectories and the final applied KOMBI bias. The accelerated paths pass through the principal intermediate basins also visited by unbiased paths with a comparable distribution. We quantify these differences using the complete transition paths in SI.3 and Fig. S3.

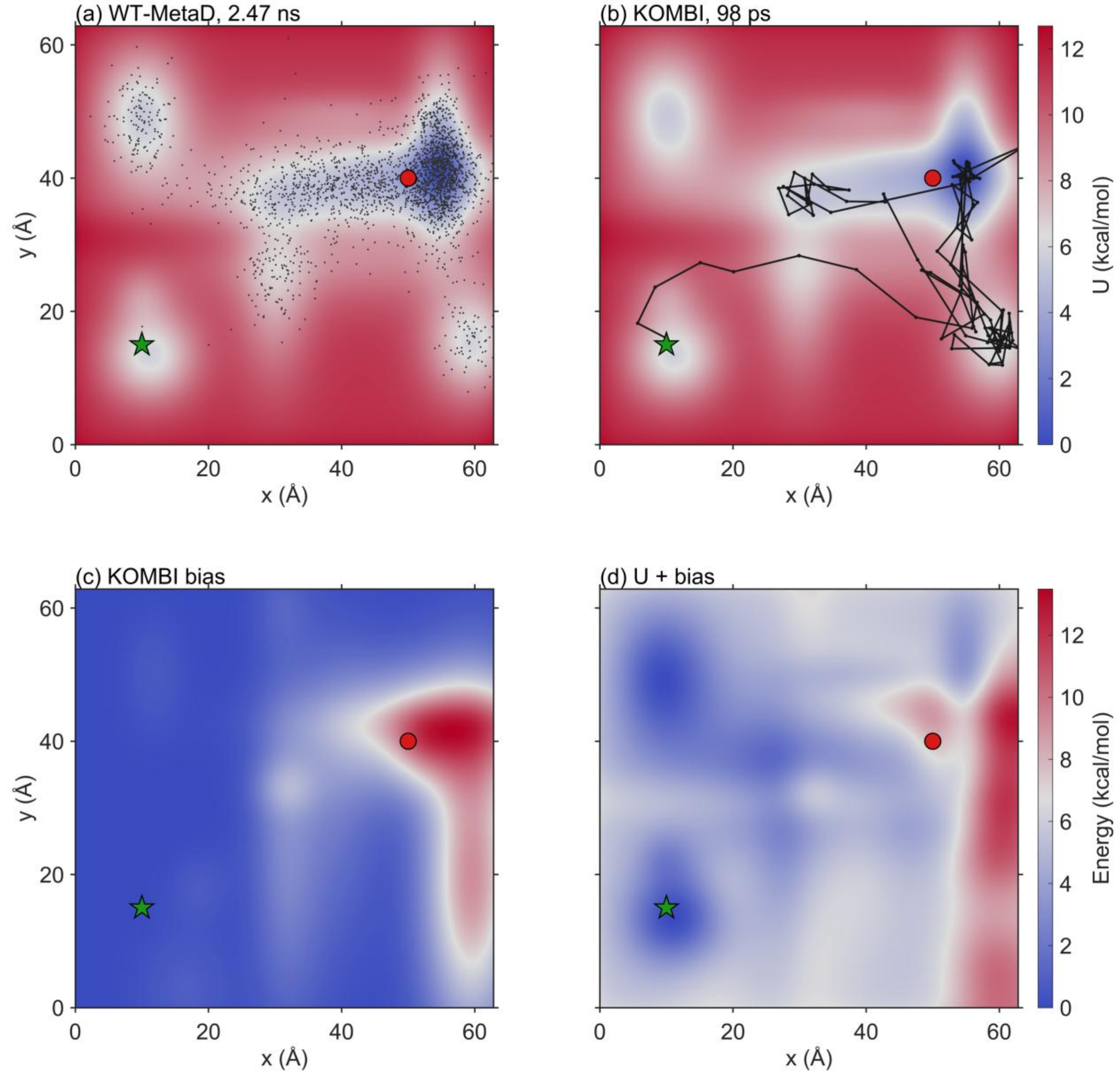

*Figure 2.* ***(a-b)*** *Trajectories for WT-MetaD* ***(a)*** *and for our KOMBI algorithm* ***(b)****, using a starting position of [50, 40] Å (red circle) and target position of [10, 15] Å (green star) on the x-y plane of the 2D free energy surface, with positions plotted every 1 ps. Trajectories were selected as the run whose first-passage time is closest to the median (WT-MetaD 2.47 ns, KOMBI 97.7 ps).* ***(c)*** *The KOMBI bias applied in the last propagation of the run in (b).* ***(d)*** *Total biased potential, defined as the original 2-D free energy surface plus the bias in (c).*

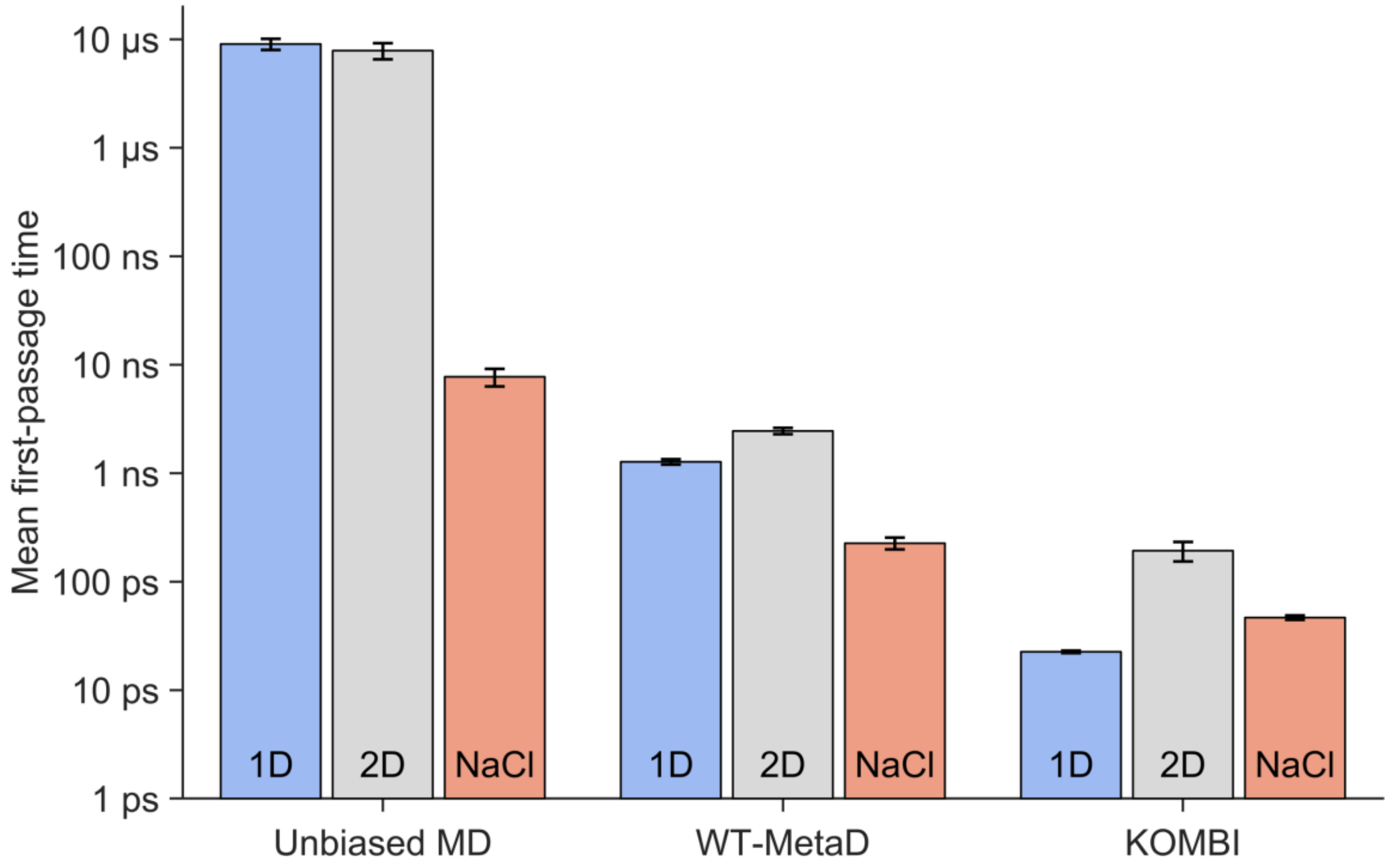

*Figure 3. Mean first-passage times for the 1D and 2D model potentials and the solvated ion-pair system (blue, grey and red bars, respectively). Bars show the mean over 30 trajectories per method and system; error bars show the standard error. All trajectories reached their targets. Unbiased times are obtained from direct simulation.*

In the NaCl system, a model ion pair with strengthened ion–ion and ion–water interactions (SI.4), the distance between the ion pair is used as our CV, with an initial distance of 2.5 Å. We define the target state as their distance reaching 8 Å following a rare event, the ion-pair dissociation, which takes 7.7 ± 1.4 ns using unbiased MD simulations. For the WT-MetaD, the system has seen significant speed-up to mean first passage times of 226 ± 28 ps. With the KOMBI algorithm, it took 46.6 ± 2.2 ps to dissociate the ion pair. The obtained MFPTs are in Fig. **3** (red bars). KOMBI was thus about 170 times faster than unbiased MD and 5 times faster than WT-MetaD.

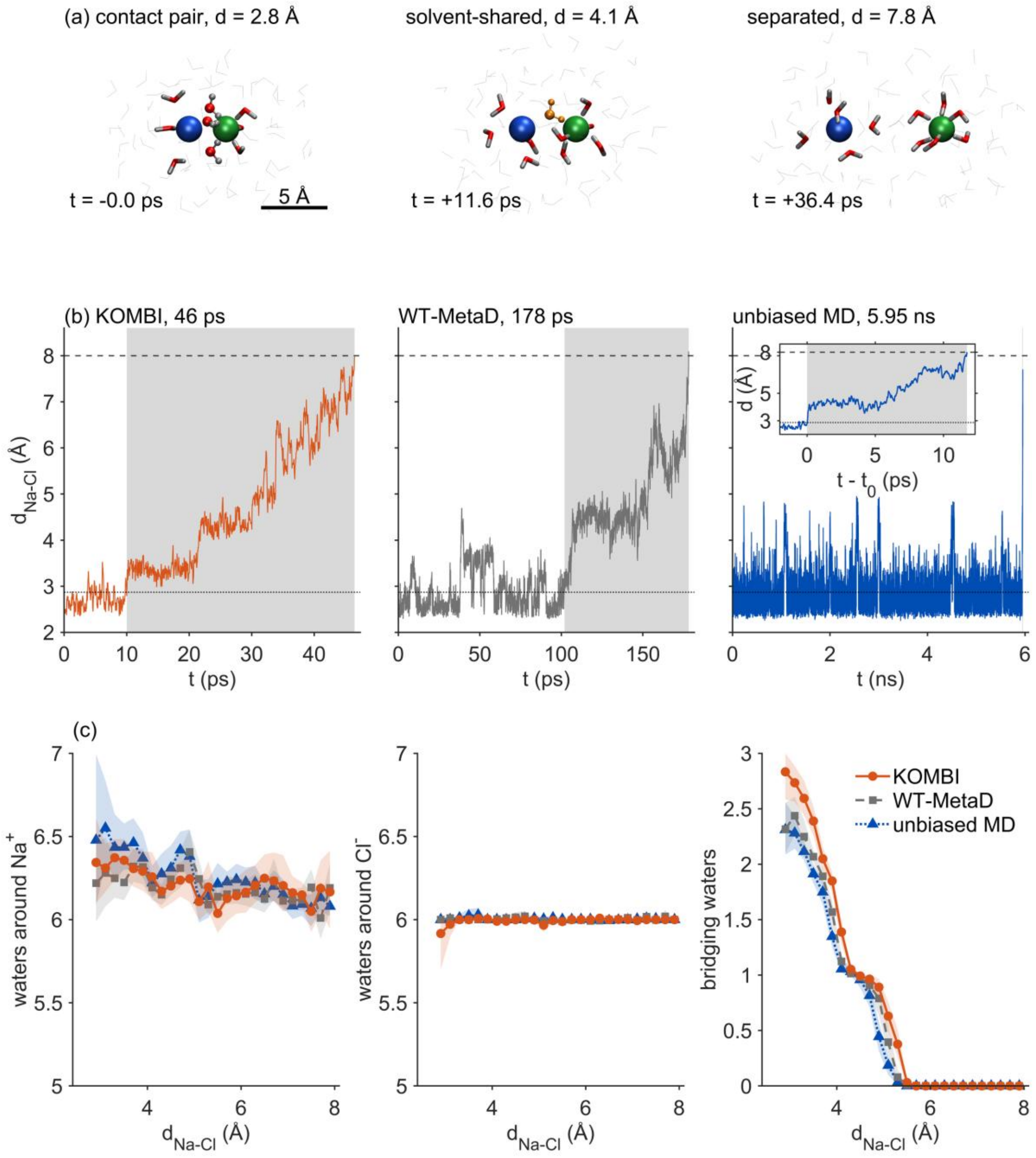

*Figure 4. Dissociation of the ion pair in the model with strengthened interactions (SI.4).* ***(a)*** *One KOMBI transition path (the run closest to the median first-passage time) at the contact pair, the solvent-shared pair and the separated pair. $Na^+$ is blue and $Cl^-$ green. Waters in the first shell of one ion are drawn as sticks, waters in both shells as balls and sticks (the one between the ions in orange), and other waters within 7 Å as lines.* ***(b)*** *Na–Cl distance against time for representative KOMBI, WT-MetaD and unbiased runs. The grey band marks the transition path. The inset enlarges the 11.7 ps unbiased transition path, with $t_0$ denoting its start. The dotted line marks the edge of the contact pair (2.87 Å) and the dashed line the target (8 Å).* ***(c)*** *Waters in the first shell of $Na^+$ and of $Cl^-$, and waters in both shells, against the Na–Cl distance along the transition paths of the 30 runs of each method. Bands are 95% bootstrap intervals over runs.*

We next applied KOMBI to phosphate ($P_i$) release from the Ras.GAP.GDP.$P_i$ complex after GTP hydrolysis, biasing the Mg–P distance. The median first-passage time to reach 8 Å separation is 0.860 ns (mean 1.166 ns, SI.5, Fig. S8a). Overlaid phosphate positions illustrate the exit region and the surrounding residues at the Ras–GAP interface (Fig. 5a). Structural analysis of 73 trajectories shows that $P_i$ leaves the $Mg^{2+}$ coordination shell after 62 ± 18 ps and subsequently remains in contact with $Q61^{Ras}$ and the catalytic arginine finger $R789^{GAP}$ for 96.7% of the recorded out-of-shell trajectory time. Q61 and R789 are conserved residues with established roles in GTP hydrolysis [39,40]. The phosphate follows a consistent exit route towards the Ras–GAP binding interface. The route extends towards GAP residue L902$^{GAP}$, which is contacted in 70 of the 73 trajectories, marking a common exit region at the Ras–GAP interface (Fig. 5b). Initial contacts and changes in $Mg^{2+}$ coordination are reported in Fig. S8b,c.

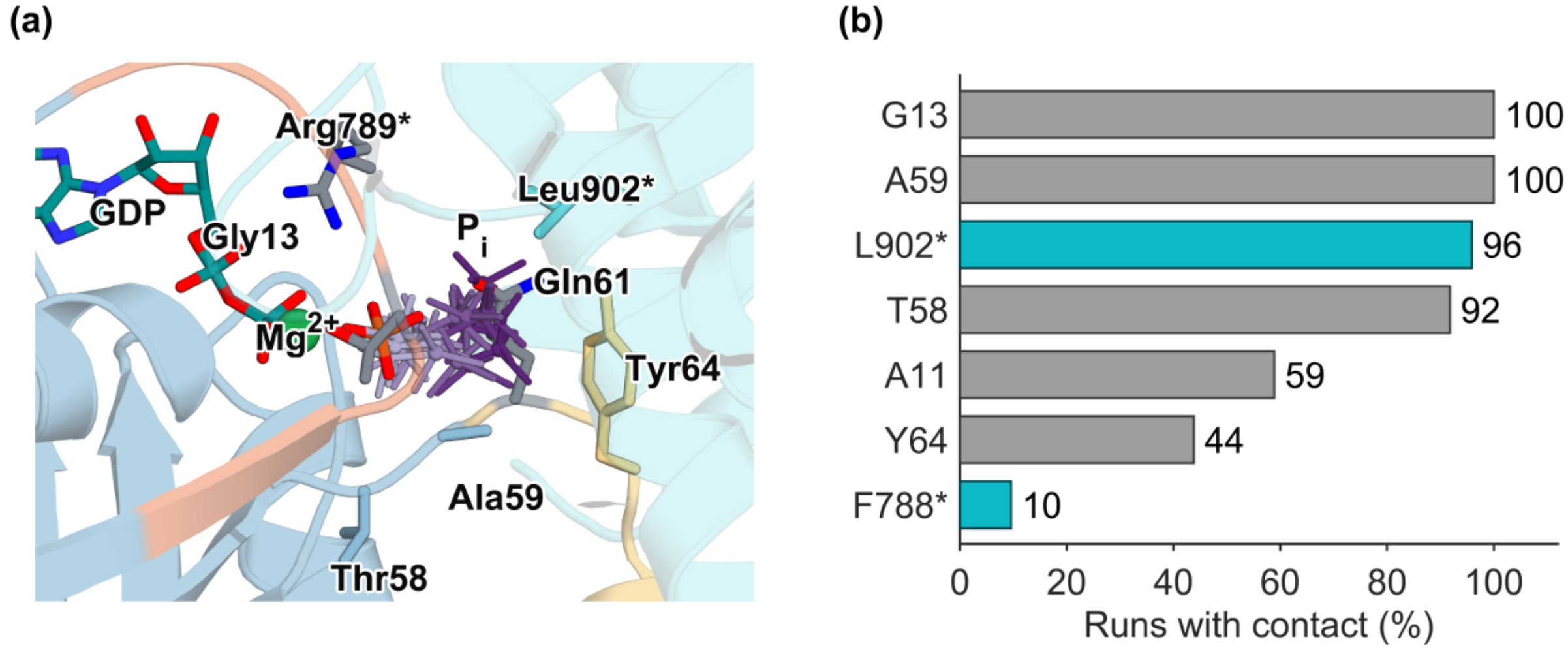

*Figure 5. New phosphate contacts formed during release from Ras.GAP.GDP.Pi. (a) Illustration of phosphate displacement towards the Ras–GAP interface, with phosphate positions (purple) from several trajectories overlaid after alignment on Ras and $Mg^{2+}$. Selected Ras and GAP residues, GDP and $Mg^{2+}$ are labelled. (b) Fractions of the 73 paths that form contacts absent from the starting structure. A contact is a Pi–protein heavy-atom distance within 3.5 Å in at least one saved frame before the first 8 Å Mg–P crossing. Only contacts present in at least 9% of paths are displayed. Gray and cyan denote Ras and GAP residues, respectively; GAP residues carry an asterisk.*

## IV. CONCLUSIONS

In this paper, we propose an MSM-based, kinetics-optimized enhanced sampling algorithm that optimizes the applied bias to minimize the MFPT to a target state. By adaptively updating the bias based on an MSM-based analysis of the discretized phase space using the DHAM method [22], our approach significantly reduces the time required to reach the target states in our test cases. These improvements are demonstrated by orders of magnitude smaller MFPTs for the 1D and 2D analytical potentials and all-atom simulations of solvated NaCl. Furthermore, our iterative bias generation method facilitates a more direct trajectory from the local starting point to the target state within the phase space, demonstrating that the KOMBI algorithm efficiently directs the system towards the global target without extensive exploration of non-essential directions in the CV space, yet providing a likely pathway.

We also provided a practical example using our novel algorithm on the Ras–GAP system. Structural analysis of the pathways for the phosphate dissociation in the post-GTP-hydrolysis state of the Ras.GAP complex was carried out, with key residues during the unbinding process identified. This underscores the utility of KOMBI as a target-directed enhanced-sampling approach that can generate physically interpretable candidate pathways for rare events in a selected CV space, while systematic validation of pathway preservation remains an important direction for future work (ESI Sections SI.3, SI.4 and SI.6). Our KOMBI algorithm provides a novel, efficient and accurate way to probe conformational changes that normally occur over long timescales using ordinary classical MD simulations. This opens up new ways to explore and facilitate rare events in the selected CV space in a highly-efficient manner, with key applications in drug discovery and materials science.

**Code and data availability.** The simulation drivers, regression tests, metadata are collected at https://github.com/rostaresearch/KOMBI.

## V. SUPPLEMENTARY MATERIAL

The supplementary material provides simulation and biasing methods; additional trajectories and pathway analyses for the one- and two-dimensional analytical systems; NaCl hydration and transition-path statistics; Ras–GAP simulation details, structural stability, phosphate first-passage times, contacts and magnesium coordination; and computational-cost information.

## VI. ACKNOWLEDGEMENT

We acknowledge discussions with Alessia Annibale, and valuable input from Dénes Berta. We acknowledge support from ERC (project 757850 BioNet), and EPSRC (grant no. EP/R013012/1). We acknowledge support from the UK Tier 2 HPC clusters and HECBioSim (http://hecbiosim.ac.uk). This work used the ARCHER2 UK National Supercomputing Service. This project made use of time on Tier 2 HPC facility JADE, funded by EPSRC (EP/P020275/1).

---

Electronic Supporting Information (ESI)

***Kinetics-Optimized Enhanced Sampling Using Mean First Passage Times***

Tiejun Wei[1], Balint Dudas[1], and Edina Rosta[1]
[1]*University College London, Department of Physics and Astronomy*
*London WC1E 6BS*
*Correspondence: e.rosta@ucl.ac.uk*

## SI.1. Common simulation and biasing methods

All enhanced sampling simulations were performed using the MD package OpenMM version 8.1 (model potentials and NaCl) and 8.2 (Ras) [1]. We designed two analytical potentials to benchmark the KOMBI approach and performed Langevin dynamics simulations on a single pseudo-atom with a mass of 12 u (atomic mass unit). The friction coefficient was 1 $ps^{-1}$ and the temperature 300 K, which gives a diffusion coefficient $k_BT/(m\gamma)$ of 2.1 × 104 $Å^2$/ns. A set of Gaussian kernels was applied to the system in addition to the original analytical potentials to create an overall potential surface for biased runs (for unbiased runs, no Gaussian functions are added). The overall potential energy (FES) applies only to the x coordinate for the 1D potential system and to the x and y coordinates for the 2D system. The 1D potential had an FES defined on the x-coordinates in the range of $[0, 2\pi]$ nm, while the 2D system was defined over both x and y coordinates in the range $[0, 2\pi]$ nm for each axis. A sigmoid potential function was applied to the boundary to restrain the system in the desired domain. A large harmonic restraint with a force constant of 1000 kJ/mol/$nm^2$ was applied on the y and z axes for the 1D and on the z axis on the 2D systems.

The Gaussian kernel used to generate the total bias $u^{\alpha}_{bias}(\xi)$ in Equation 4 is defined as below in the 1D/2D potential energy surface:

$$G_i^{\alpha}(\mu_i^{\alpha}, A_i^{\alpha}, \sigma_i^{\alpha}; \xi) = A_i^{\alpha} \exp\left[-\frac{(\xi-\mu_i^{\alpha})^2}{2(\sigma_i^{\alpha})^2}\right] \quad (S.1)$$

$$G_i^{\alpha}(A_i^{\alpha}, \mu_{xi}^{\alpha}, \mu_{yi}^{\alpha}, \sigma_{xi}^{\alpha}, \sigma_{yi}^{\alpha}; \xi) = A_i^{\alpha} \exp\left[-\frac{(x-\mu_{xi}^{\alpha})^2}{2(\sigma_{xi}^{\alpha})^2} - \frac{(y-\mu_{yi}^{\alpha})^2}{2(\sigma_{yi}^{\alpha})^2}\right] \quad (S.2)$$

Where the $A^{\alpha}, \mu^{\alpha}, \sigma^{\alpha}$ are the amplitude, mean and width of the Gaussian Kernel, and ξ is the CV space, where in 1D ξ is the x-coordinate of the pseudo-atom, in 2D ξ is the x, y coordinates of the pseudo-atom.
We use the best randomly obtained initial guess from 1000 trials for the Gaussian kernel parameters. This is subsequently optimized using the L-BFGS-B [2] optimizer with SciPy [3], with boundaries of the constraint positions ($\mu_{xi}^{\alpha}$) corresponding to the upper and lower limits of the selected reaction coordinates. Each Gaussian amplitude, $A_i^{\alpha}$, was bounded between 0.1 kcal/mol and 1.2 kcal/mol (1D, 20 Gaussians), 2 kcal/mol (2D, 20 Gaussians; NaCl, 10 Gaussians) or 5 kcal/mol (Ras, 10 Gaussians). The total bias was not bounded.
For the MSM/DHAM analysis, the selected CV space was discretized using uniform bins along each coordinate, with the bin width chosen to resolve minimal coordinate fluctuations observed during MD simulations: approximately 0.07 Å for the NaCl distance over the 2–9 Å range, approximately 0.07 Å for the P–Mg distance over the 1–8 Å range, 100 bins on [0, 2π] for the 1-D model potential, and a 20 × 20 grid on $[0, 2\pi]^2$ for the 2-D model potential. Transition counts were accumulated from the CV recorded at every MD step, at a lag of one step (τ = 2 fs), in all systems. The local target state was defined as the visited bin closest to the final target in CV space, and DHAM was used to combine transition information from trajectories generated under different bias potentials using the Rosta–Hummer reweighting relation, with the unbiased transition matrix obtained in a single DHAM step and normalized to conserve probability. MFPTs are evaluated by an absorbing-state solve or numerically stable state reduction. When the current state is also the selected local target, the objective is the first-return time after at least one model step, rather than the zero-time hitting time.
For the WT-MetaD comparison, the following parameters were used. Gaussian widths were 0.02 nm for NaCl and 0.2 nm for the 1-D and 2-D Langevin systems; Gaussian heights were 1.5 kcal/mol for NaCl and 3 kJ/mol for the Langevin systems; Gaussians were deposited every 5000 MD steps (10 ps for a 2 fs time step); and the bias factor was 5.0.

Unbiased first-passage times on the 1D and 2D model potentials were obtained from 30 direct Langevin simulations for each potential. Because these unbiased runs require up to 1.3 × 1010 steps, we integrated the same Langevin update as OpenMM's LangevinIntegrator (m = 12 u, γ = 1 ps−1, 2 fs time step, 300 K) with a numba implementation, which we checked against OpenMM on short transitions (mean first-passage times within 6%). Each run started from the energy minimum nearest to the KOMBI starting position and ended at the first arrival within 1 Å of the KOMBI target. The results are reported separately in SI.2 and SI.3.

**SI.2. One-dimensional analytical system**

The separate exact-potential illustration in Fig. 1 uses 100 states and 30 Gaussians of fixed height 2 kcal/mol. Centers range from 0 to 99 and widths from 0.5 to 50 in state-index units. Eight seeded L-BFGS-B optimizations are performed per starting state using the exact birth–death MFPT and its analytical gradient. The rate prefactor is one per model time unit. The optimized bias in these examples are not the adaptive MD protocol used for Fig. 3, but obtained using the exact potential energy.

For the reference unbiased simulations on the 1D potential, the 30 first-passage times had a mean of 9.0 ± 1.1 μs. A representative trajectory is shown in Fig. S1b.

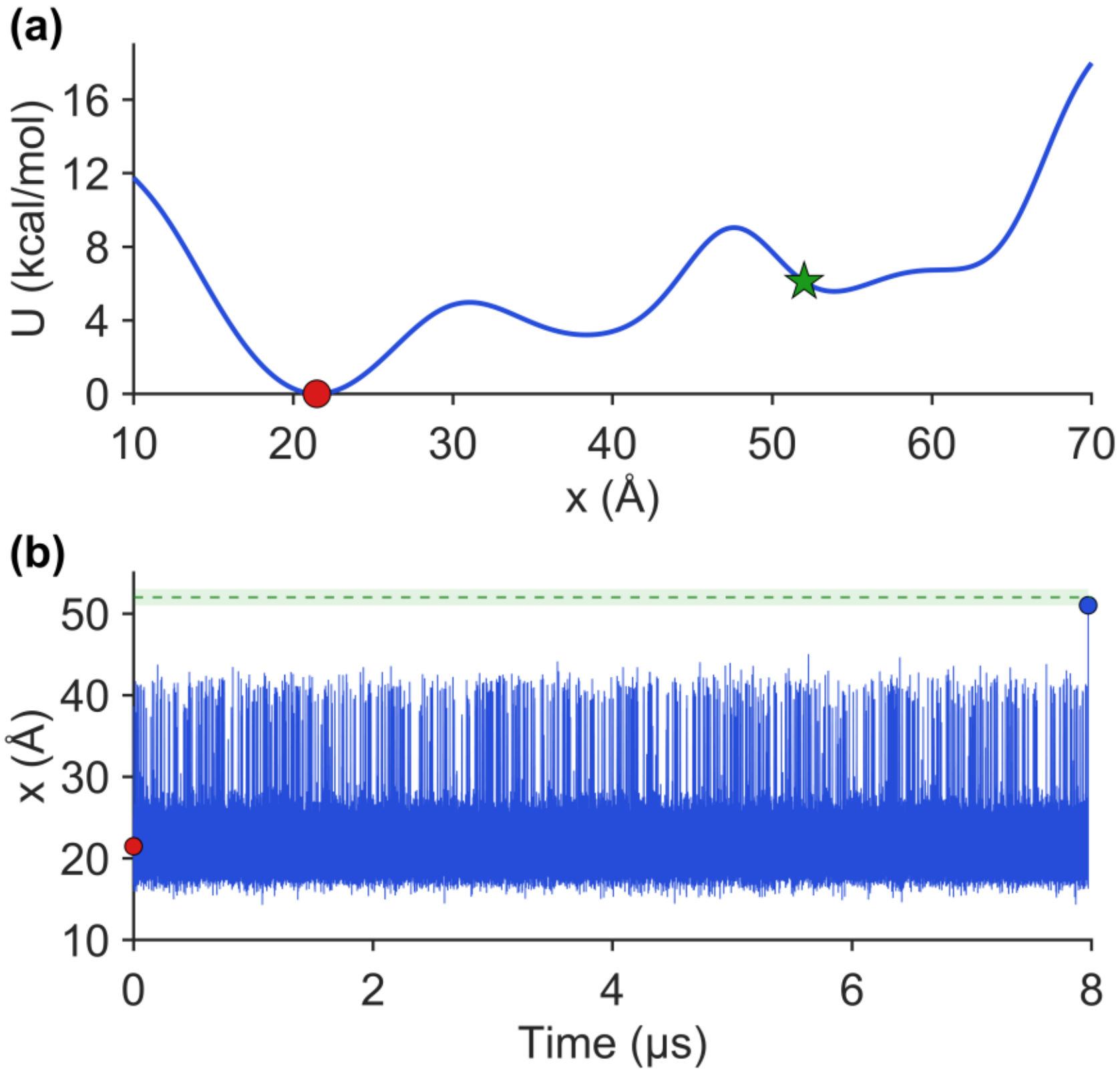

***Figure S1.*** *Representative unbiased Langevin trajectory on the 1D model potential. (a) Potential, with the starting minimum (red circle) and target at 52 Å (green star). (b) Position versus time for the run whose first-passage time is closest to the median of 30 runs (7.97 μs). The green band marks the target region within 1 Å of 52 Å; the dashed line marks its centre. Samples recorded every 10 ps are supplemented by 1 ps samples during the final 100 ps and the exact first-arrival frame.*

*Alt text: The one-dimensional potential marks the initial basin near 22 angstroms and a target near 52 angstroms. A representative unbiased Langevin trajectory fluctuates in the initial region for almost 8 microseconds before first reaching the target.*

## SI.3. Two-dimensional analytical system

For the 2D potential, the 30 unbiased first-passage times had a mean of 7.9 ± 1.3 μs. Representative trajectories are shown in Fig. S2.

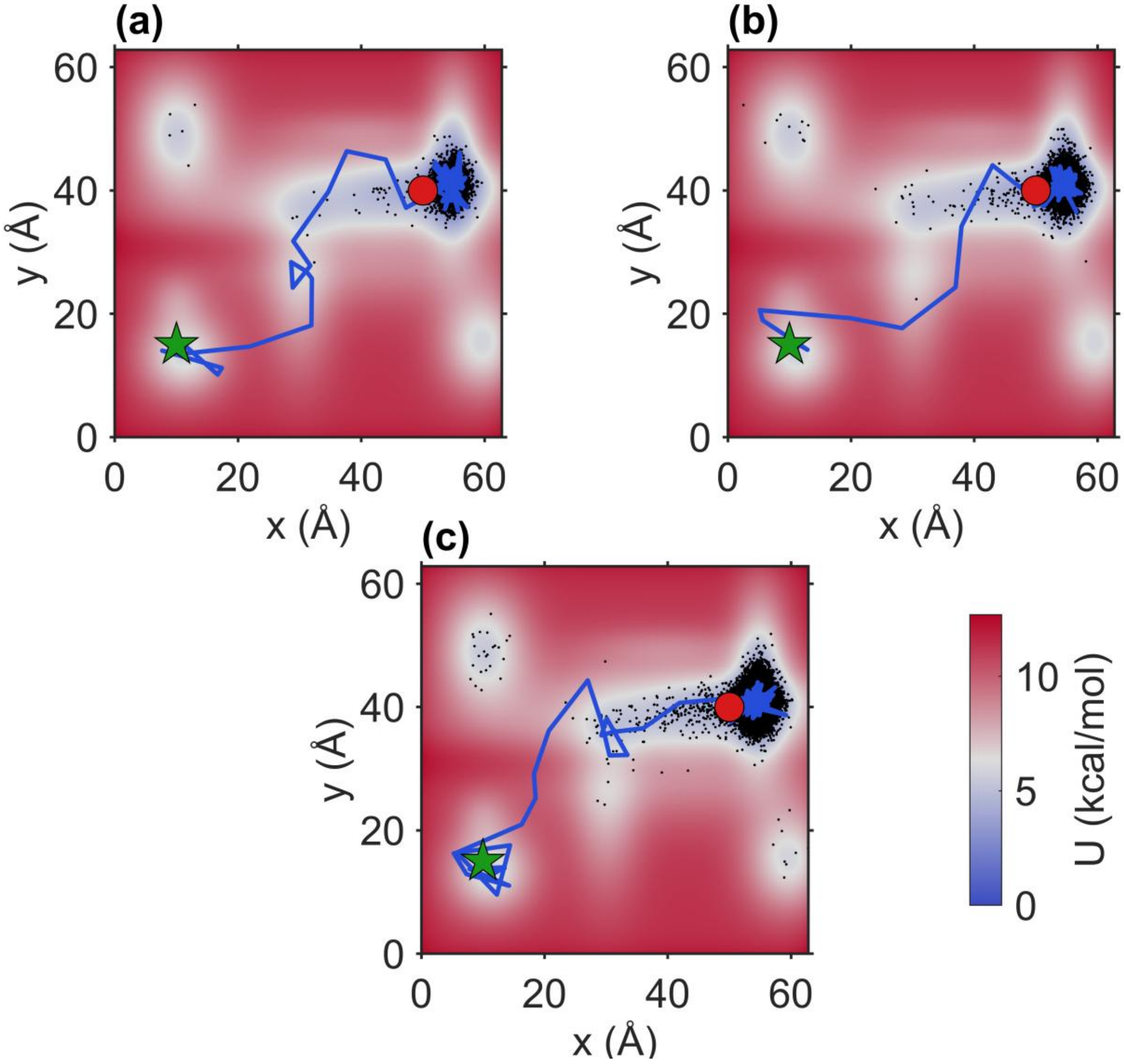

***Figure S2.*** *Unbiased Langevin trajectories on the 2D model potential. Panels (a–c) show the runs closest to the 25th, 50th and 75th percentiles of 30 runs (1.91, 4.42 and 12.82 μs). Positions are shown every 200 ps (dots), with the last 100 ps every 1 ps (blue line). Red circles mark the starting positions and green stars the target at [10, 15] Å.*

*Alt text: Three potential-energy heat maps show representative unbiased two-dimensional Langevin trajectories near the 25th, 50th, and 75th percentiles of first-passage times. All begin near the same upper-right basin and reach the lower-left target through intermediate regions.*

We compared the transition paths of the 30 KOMBI runs and the 30 unbiased runs of Fig. 3 (Fig. S3). The KOMBI runs saved the positions at every step; for the unbiased runs, the positions at every step were recovered by repeating each run from its random seed, which reproduces it exactly. A transition path starts when the particle last leaves the start basin (more than 3.5 Å from its minimum) and ends when it reaches the target. A path visits one of the other four basins when it comes within 2 Å of its minimum.

Both sets visit the principal intermediate basins, with comparable visit frequencies. Basins 2 and 3 are visited by 70% and 63% of the directed KOMBI paths, compared with 73% and 50% of unbiased paths. Basin 1 is visited by 37% and 17%, respectively, and basin 4 by one of 30 KOMBI paths and none of the unbiased paths. The median transition-path duration is 38.2 ps for KOMBI and 22.3 ps for unbiased dynamics.

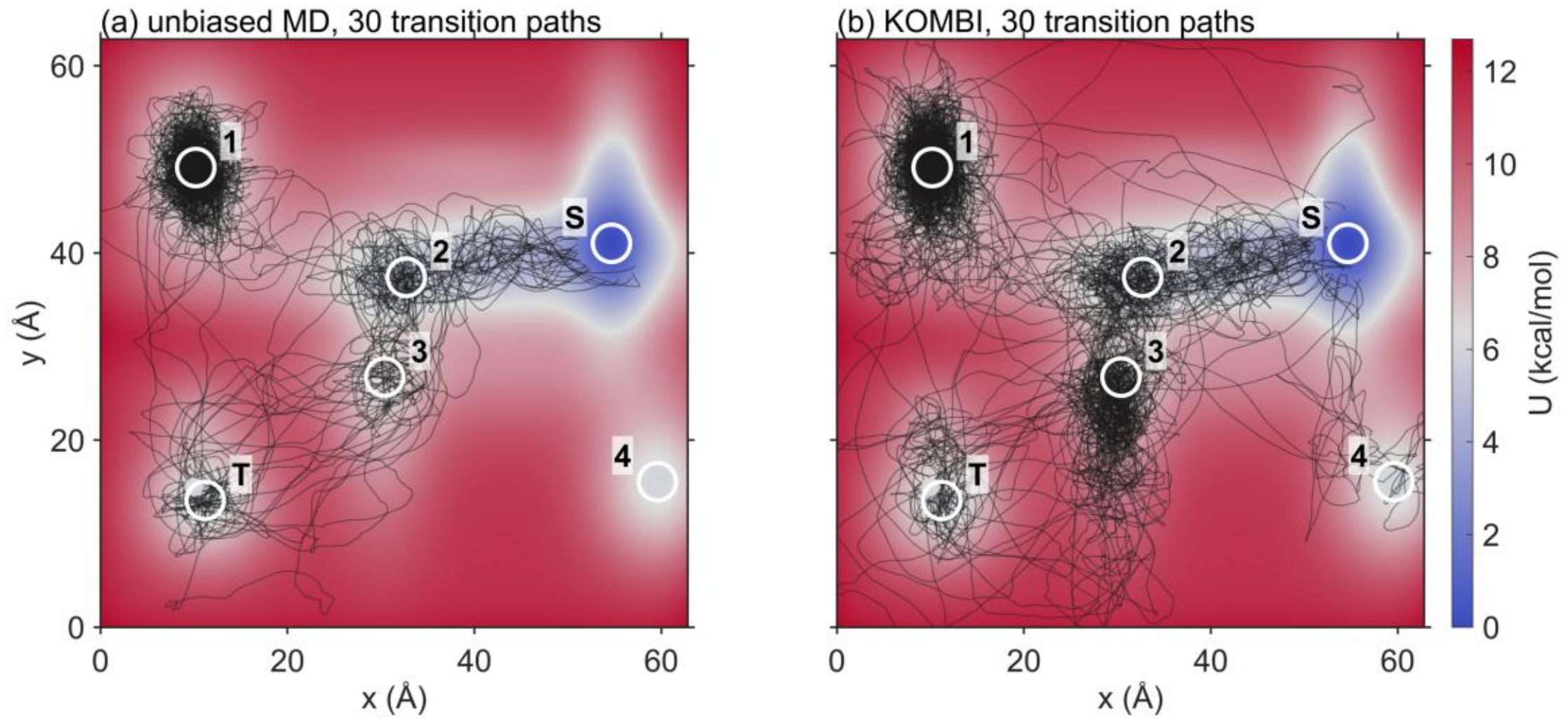

**Figure S3.** Transition paths on the 2D potential. (a) The 30 unbiased runs and (b) the 30 KOMBI runs of Fig. 3, each from the last exit from the start basin to the target, drawn on the potential energy U. Circles mark the basin minima: S, start; T, target; 1–4, the other basins.

Alt text: Two heat maps overlay complete transition paths from 30 unbiased trajectories and 30 KOMBI trajectories on the same two-dimensional potential. Both sets commonly visit intermediate basins two and three, while basin one is visited more often during KOMBI trajectories.

## SI.4. NaCl simulations.

A single $Na^+$-$Cl^-$ ion pair was solvated with 566 TIP3P water molecules [4]. The ion parameters were taken from the parameter set by Roux and co-workers [5]. We modified the Lennard-Jones interaction to raise the energy barrier separating the pair. This was done by modifying the ε (the depth of the potential well) and $R_{min}$ (the distance at the minimum of the pair potential) values in the NBFIX term of CLA-SOD from -0.0839 kcal/mol to -6.0 kcal/mol and from 3.7310 Å to 2.5 Å [6], respectively. In addition, the Lennard-Jones parameters of both ions were set to ε = −5.0 kcal/mol and $R_{min}/2$ = 1.0 Å (standard CHARMM values: −0.0469 kcal/mol and 1.411 Å for $Na^+$, −0.150 kcal/mol and 2.27 Å for $Cl^-$). Together, these changes strengthen the ion–ion and ion–water interactions, so that the system serves as a ligand-unbinding-like example with a sizeable barrier. These can induce structural changes, for example, the first Cl–O peak of the radial distribution function lies at 2.3 Å, against about 3.1 Å with the standard parameters. The system was minimized for 1000 steps before production to avoid separating the ion pair during preparation.

The ion pair and the 566 water molecules were simulated in a cubic periodic box with an edge of 25.6 Å, set by 1 ns of NPT simulation at 300 K and 1 bar (density 1.017 g $cm^{-3}$). Electrostatics were computed with particle-mesh Ewald and a 1.0 nm real-space cutoff, which was also used for the Lennard-Jones interactions. Langevin dynamics at 300 K used a friction coefficient of 10 $ps^{-1}$ and a 2 fs time step, with bonds to hydrogen constrained. All simulations started from the contact pair (2.5 Å) and ended when the Na–Cl distance first reached 8 Å. The coordinates of all atoms were saved every 0.1 ps (for unbiased MD, over the last 1 ns before the first passage).

**Ion pair.** In this model, the first peaks of the ion–water radial distribution functions are at 2.41 Å (Na–O) and 2.29 Å (Cl–O), and the first minima at 3.19 Å and 2.83 Å, which we use as the first-shell cutoffs (Fig. S4a). Each ion has about six water molecules in its first shell, both at contact and when the ions are apart. The transformed radial occupancy has a minimum near 2.52 Å (Fig. S4b). We retain 2.87 Å as an operational contact boundary for all three methods. A transition path runs from its last exit from this region to first arrival at 8 Å; this geometric boundary is not interpreted as a committor surface. A bridging water is a water molecule in the first shells of both ions; it sits between the ions when the Na–O–Cl angle is at least 120°.

The three methods give the same sequence of events (Fig. 4). From the contact pair, a water molecule enters between the ions at d = 4.1–4.2 Å (medians of the three methods; Fig. S5b), in every run; the bridge then breaks and the ions separate. Along the path, each ion keeps about six water molecules in its first shell (Fig. 4c). The transition paths are slightly shorter with the biases (medians 36 ps for KOMBI, 42 ps for WT-MetaD and 53 ps for unbiased MD; Fig. S5a). The clearest difference is when the bridging water leaves (Fig. S6). The bridge has broken in half of the frames at d = 5.20 Å in the KOMBI paths, at 5.05 Å with WT-MetaD and at 4.87 Å in unbiased MD, as shown by the run-weighted occupancy curves. This is not because the KOMBI paths cross this region faster: they spend as long at d = 4.6–5.6 Å as the unbiased paths (medians 6.7 and 6.6 ps).

The solvent comparison resolves differences not apparent from the ion separation alone. At d = 4.8–5.0 Å, bridging waters are more frequent in unbiased excursions that return to the contact pair than in reactive paths. This association motivates including solvent variables when describing the transition.

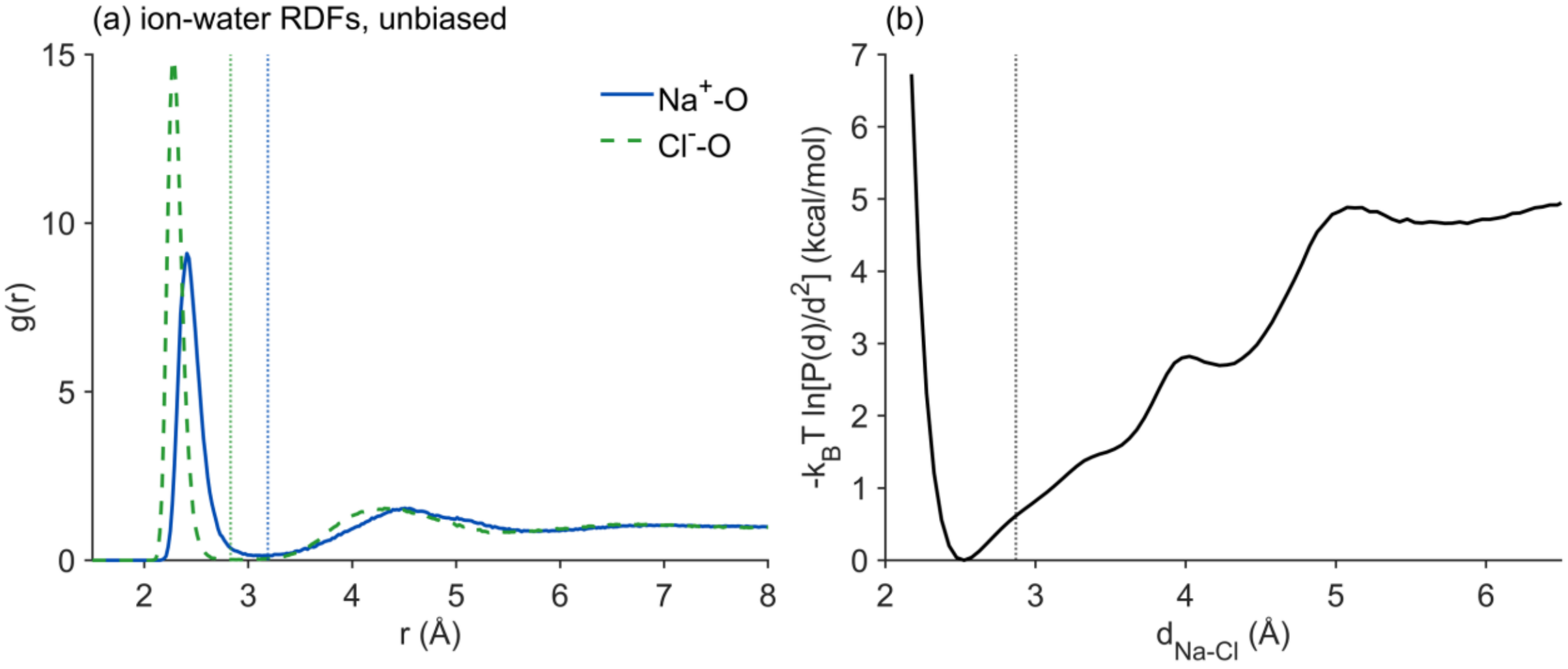

Figure S4. Hydration and radial occupancy of the ion-pair model from unbiased first-passage trajectories. (a) Na–O and Cl–O radial distribution functions; dotted lines mark the first-shell cutoffs. (b) The transformed occupancy $-k_BT \ln[P(d)/d^2]$, with the operational contact boundary at 2.87 Å.

Alt text: Two plots characterize the modified NaCl ion pair in water. Radial distribution functions identify first-shell water distances around sodium and chloride; a transformed radial occupancy profile shows the ion-separation distribution and the operational contact-pair boundary of 2.87 angstroms.

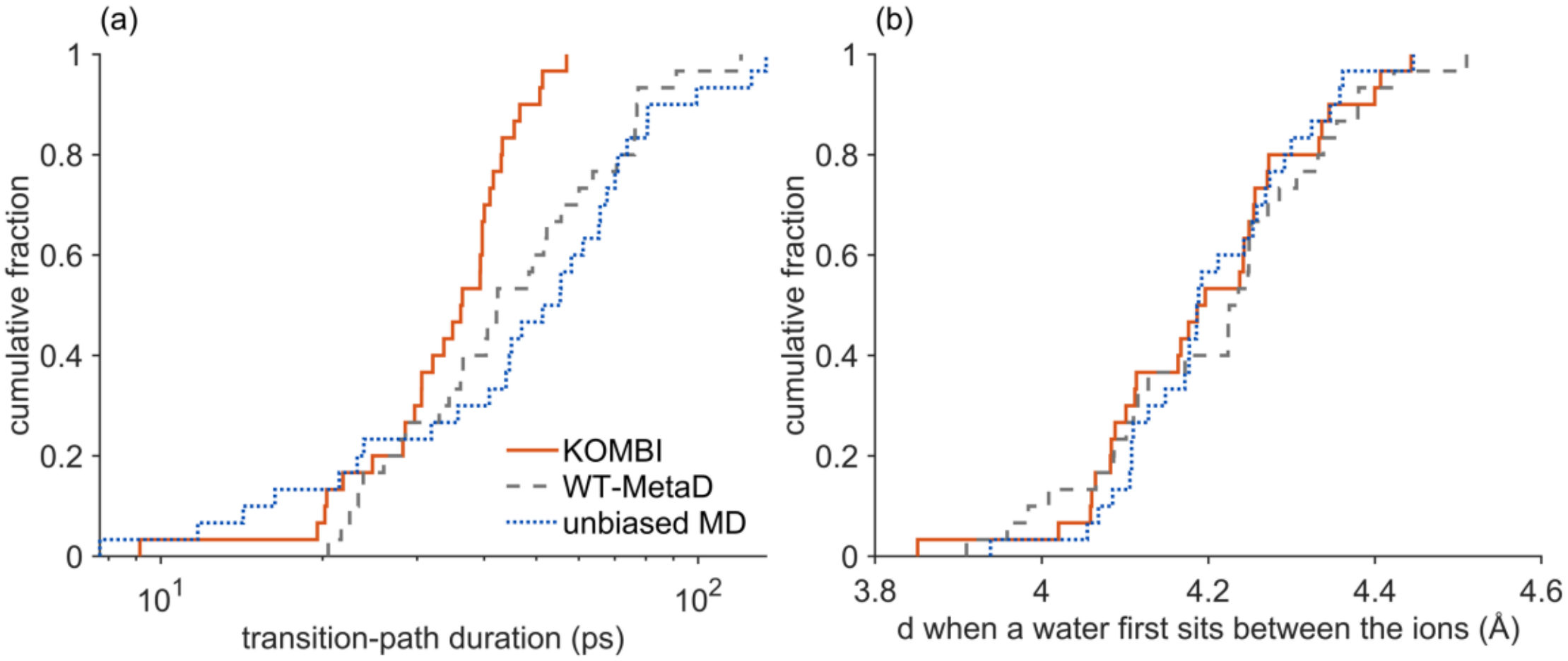

**Figure S5.** Cumulative distributions over the 30 runs of each method of (a) the transition-path duration and (b) the Na–Cl distance at which a water molecule first sits between the ions.

Alt text: Two cumulative distribution plots compare 30 trajectories per sampling method. The first shows transition-path durations, with shorter median durations for the biased methods. The second shows the NaCl distance where a water molecule first bridges the ions, clustering around 4.1 to 4.2 angstroms.

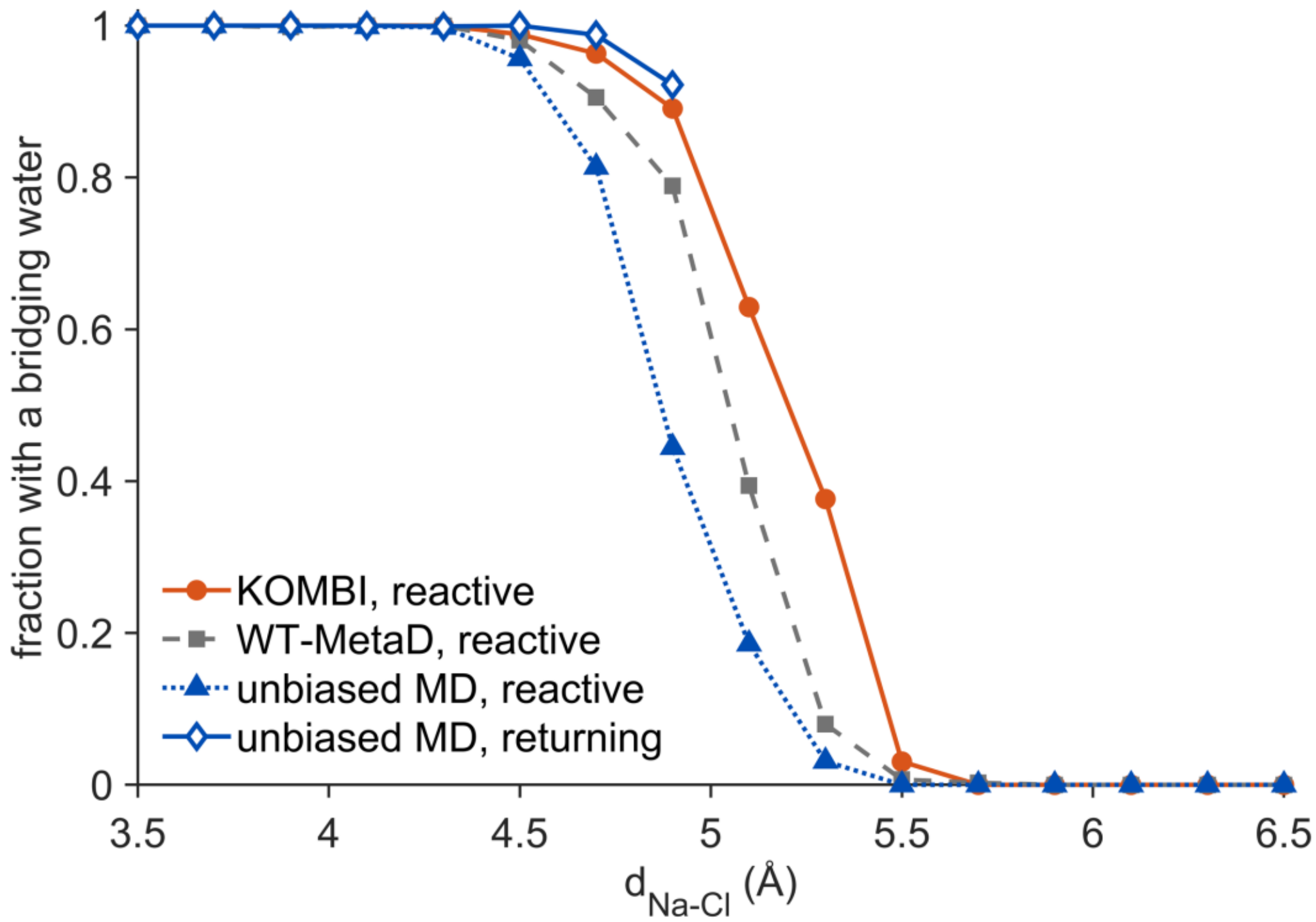

**Figure S6.** Fraction of frames containing a bridging water molecule as a function of Na–Cl distance. Results are shown for transition-path frames from all three methods (averaged across runs) and, for unbiased MD, for frames from the preceding 1 ns that lie outside the transition path, including excursions from the contact ion pair that return to it.

Alt text: Line plots show bridging-water occupancy as sodium and chloride separate. Reactive trajectories from KOMBI, well-tempered metadynamics, and unbiased dynamics lose the bridging water as separation increases. Returning unbiased excursions retain bridging water more often near 4.8 to 5.0 angstroms.

## SI.5. Ras–GAP phosphate release

The $GDP^-$, $H_2PO_4^-$, and $Mg^{2+}$-bound Ras.GAP complex system was built using the product state of our previous QM/MM simulations for the Ras.GAP complex [7]. A rectangular box of TIP3P [4] water molecules with 15 Å in all directions from the protein surface was generated with CHARMM-GUI [8,9], we set a KCl concentration to 0.15 M. The solvated system contains 30,630 TIP3 water molecules [4], 102 $K^+$, and 86 $Cl^-$ ions. The solvated system was energy minimized using the steepest descent method for 10,000 steps, which was followed by 10 ns long NVT equilibration and, subsequently, 10 ns long NPT production runs at 303.15 K temperature. Simulations were performed with NAMD3.0 [10] using the additive all-atom CHARMM force field C36m [11]. For constant temperature control, Langevin dynamics was used with a damping coefficient of 1 $ps^{-1}$. The constant pressure was achieved by using the Nose-Hoover method with a piston oscillation period of 50 fs, and a piston oscillation decay time of 25 fs. The integration time step was set to 2 fs. The particle mesh Ewald (PME) method was used to calculate the electrostatic interactions with a maximum grid spacing of 1 Å and interpolation order of 6. Van der Waals interactions were reduced to zero by ‘switch’ truncation operating between 10.0 Å and 12.0 Å. The last frame of the unbiased production run was used in subsequent enhanced sampling simulations using the KOMBI algorithm.

The $P_i$ unbinding simulations started from the equilibrated periodic Ras.GAP.GDP.$P_i$ structure and biased the minimum-image distance between $Mg^{2+}$ and the $P_i$ phosphorus. We generated 73 completed paths, their first-passage times are shown in Fig. S8a. Their mean, median and range of arrival times are 1.166 ns, 0.860 ns and 0.090–4.326 ns, respectively.

The KOMBI simulations used OpenMM with periodic boundaries, particle-mesh Ewald electrostatics, a 1.2 nm real-space cutoff and a 1.0–1.2 nm Lennard-Jones switching interval. Langevin dynamics at 300 K used a friction coefficient of 1 $ps^{-1}$, a 2 fs time step and constraints on bonds to hydrogen. The initial periodic state was obtained from the equilibration described below. The bias was a sum of ten Gaussians in the Mg–P distance, with amplitudes of 0.1–5 kcal $mol^{-1}$, centers in 1–8 Å and widths of 0.7–2.0 Å. These bounds apply per Gaussian, not to the sum. Each 20 ps propagation contributed directed counts at a lag of one integration step. The initial small bias had amplitude 0.01 kcal $mol^{-1}$ per Gaussian. Applied bias energies were checked against their analytical Gaussian values during each run. Sampling and analysis ended at the first minimum-image Mg–P distance of at least 8 Å.

For the periodic starting-state preparation and stability check, the system was simulated in OpenMM 8.2, with particle-mesh Ewald electrostatics, a 1.2 nm cutoff with a 1.0–1.2 nm switch for the Lennard-Jones interactions and an initial cubic box of 10.0 nm. After 1.6 ns of equilibration (restrained NVT and NPT, then unrestrained NPT at 1 bar), three independent 10 ns NPT runs stayed close to the starting structure (Fig. S7): over the last 5 ns, the Cα RMSD was 0.83–0.89 Å for Ras and about 1.55 Å for GAP, and Pi, Ser17, Thr35 and two water molecules coordinated the $Mg^{2+}$ ion in at least 99.9% of the frames.

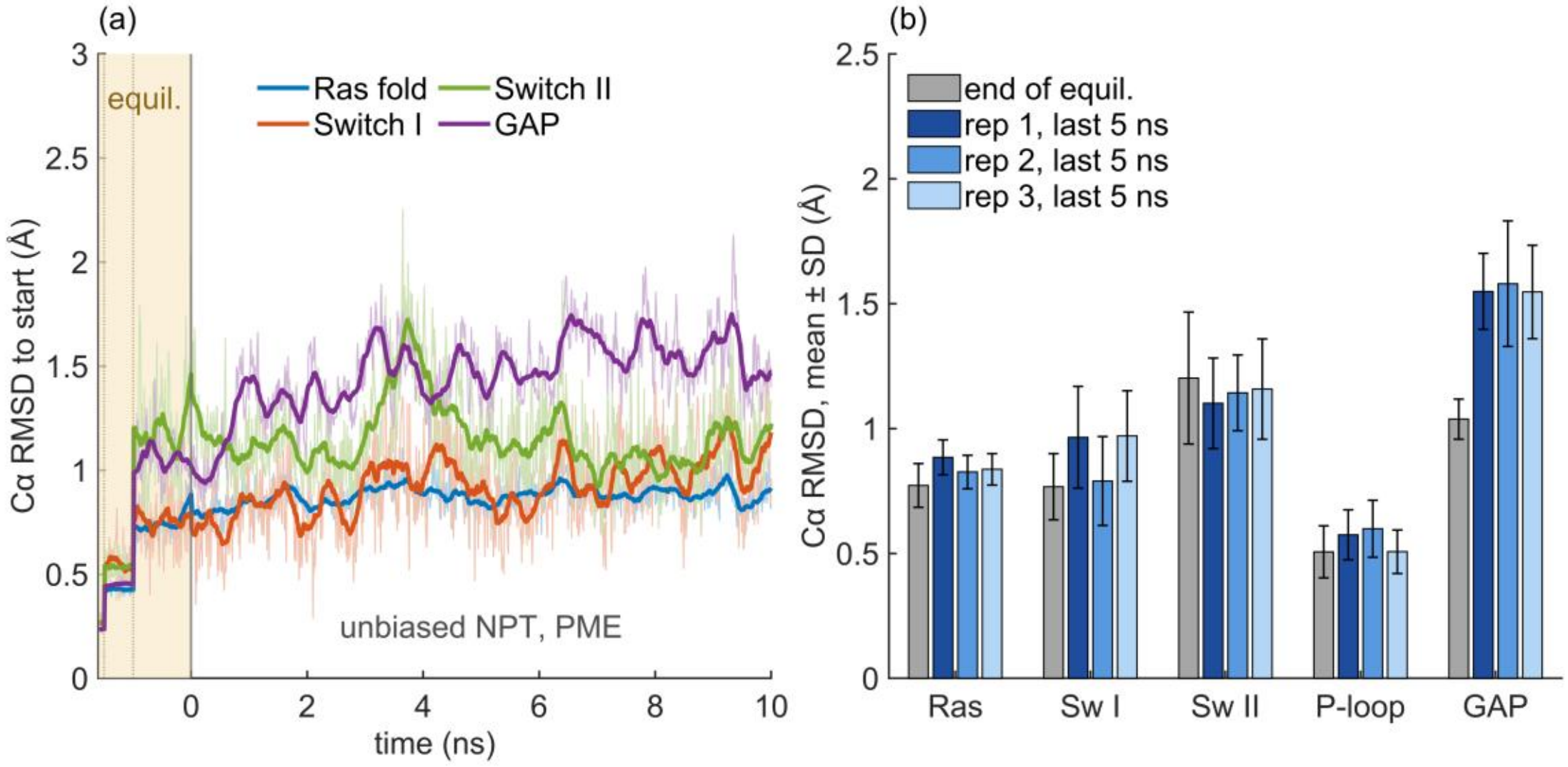

**Figure S7.** Ras.GAP.GDP.$Mg^{2+}$.$P_i$ with periodic boundaries (unbiased NPT, particle-mesh Ewald). (a) Cα RMSD to the KOMBI starting structure in replicate 1 for the Ras fold, switch I, switch II and GAP; the shaded region is the equilibration. Thin lines: every 10 ps; thick lines: 250 ps moving average. (b) Mean ± SD over the last 5 ns of each of the three replicates, with the end of the equilibration for reference.

Alt text: Protein backbone root-mean-square deviations are plotted during equilibration and three unbiased Ras-GAP simulations. The Ras fold, switch regions, and GAP remain near their prepared structure, with mean deviations over the final five nanoseconds lower for Ras than GAP.

In all 73 completed paths, $P_i$ left the $Mg^{2+}$ coordination sphere (no $P_i$ oxygen within 3.0 Å for ten consecutive saved 1 ps frames), after 62 ± 18 ps. A second nonbridging $P_\beta$ oxygen coordinated $Mg^{2+}$ at some point in every path and for at least ten consecutive frames in 70 paths. A $P_\alpha$ oxygen also coordinated $Mg^{2+}$ in 14 paths, for at least ten frames in seven. No third or fourth coordinating water was observed before first passage. Water occupancies count coordinating molecules, irrespective of their identities.

$T35^{Ras}$ left the $Mg^{2+}$ shell for at least ten consecutive saved frames in seven of the 73 completed paths, with a median first release time of 0.352 ns. $D57^{Ras}$ entered and remained coordinated for at least ten frames in two paths. The endpoint coordination fractions in Fig. S8c use the last saved 1 ps frame at or before the exact per-step 8 Å crossing. Protein contacts in Fig. 5b and Fig. S8b require a $P_i$ heavy atom within 3.5 Å of a residue heavy atom in at least one such frame. Contacts are classified as initial or newly formed using the equilibrated starting coordinates and the same 3.5 Å cutoff. GAP numbering follows the original structure mapping (topology residue number plus 717).

The $Mg^{2+}$ coordination shell also rearranges during $P_i$ departure (Fig. S8c). At the last saved frame before the 8 Å crossing, $S17^{Ras}$ and one nonbridging GDP $P_\beta$ oxygen coordinate $Mg^{2+}$ in all 73 paths; a second $P_\beta$ oxygen is present in 71 paths (97.3%), and $T35^{Ras}$ remains coordinated in 67 (91.8%). A $P_\alpha$ oxygen is coordinated in six paths (8.2%) and $D57^{Ras}$ in two (2.7%). No third coordinating water is observed before the endpoint.

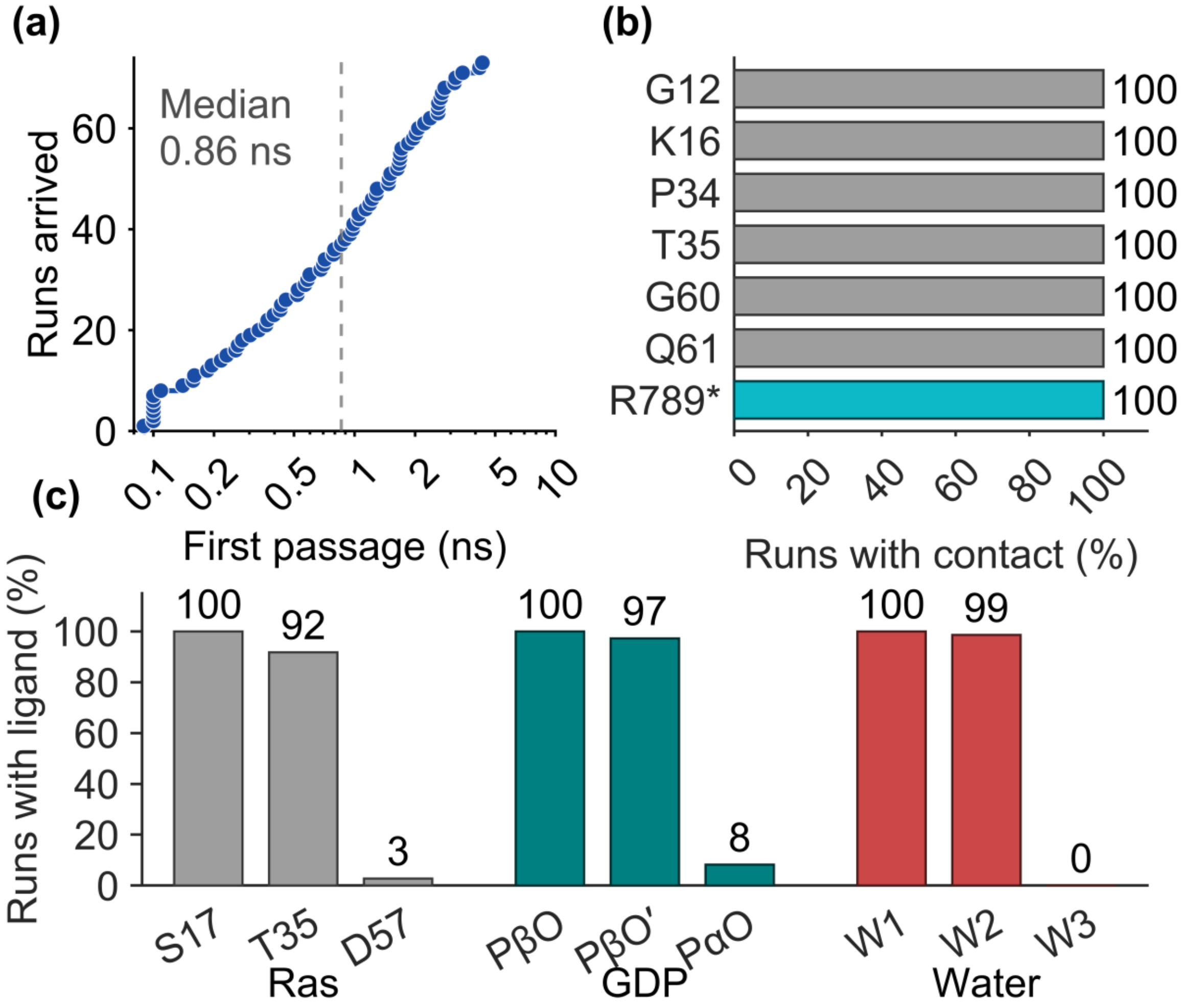

*Figure S8. First-passage times, initial phosphate contacts and $Mg^{2+}$ coordination for the 73 Ras trajectories with 20 ps bias updates. (a) Cumulative number of trajectories reaching an Mg–P separation of 8 Å; the dashed line marks the median first-passage time of 0.860 ns. (b) Fractions of paths with Pi–protein contacts already present in the starting structure. Contacts use a 3.5 Å heavy-atom cutoff; only residues contacted in at least 9% of paths are displayed. Gray and cyan denote Ras and GAP residues; GAP residues carry an asterisk. (c) $Mg^{2+}$ ligands within 3.0 Å at the last saved 1 ps frame at or before the first 8 Å Mg–P crossing. W1, W2 and W3 denote at least one, two and three coordinating waters, not conserved water identities. PβO and PβO′ denote the first and second coordinating GDP beta-phosphate oxygens; PαO denotes an alpha-phosphate oxygen.*

Alt text: A cumulative curve shows 73 Ras-GAP phosphate trajectories reaching an 8 angstrom magnesium-phosphorus distance, with a median first-passage time of 0.860 nanoseconds. Bar charts summarize initial phosphate-residue contacts and magnesium coordination at the endpoint, including Ser17 and GDP phosphate oxygens.

**SI.6. Computational cost and current limitations**

The computational cost of KOMBI is dominated by the MD propagation steps, particularly for the all-atom systems. The additional cost of constructing the MSM/DHAM transition matrix and optimizing the Gaussian bias parameters is comparatively small for the low-dimensional CV spaces considered here, because these operations are performed on the discretized CV-space representation rather than on the full atomistic configuration space. In the present implementation, the Gaussian parameters were optimized numerically; future implementations using analytical gradients are expected to further reduce this optimization overhead.

The current implementation is intended for target-directed enhanced sampling in a selected CV space. As with other CV-based enhanced-sampling methods, its performance depends on the quality of the chosen

CVs and on the availability of a meaningful target state. KOMBI is therefore best suited to cases where the initial and target states can be defined in advance and where the selected CVs capture the slow transition of interest. The method may be less effective when important slow degrees of freedom are missing from the CVs, when the target state is not known, or when many competing pathways exist in a highly rugged landscape. In principle, extension to higher-dimensional CV spaces is possible, but the discretized MSM representation can scale rapidly with dimensionality. Local target selection and adaptive exploration can partly mitigate this issue in practice, and more scalable implementations will be addressed in future work.

The biased trajectories generated by KOMBI should be interpreted as accelerated, target-reaching trajectories in the chosen CV space. In the Ras.GAP application, the repeated simulations identify recurring structural features along phosphate dissociation, but a systematic validation of pathway preservation and comparison with unbiased transition-path ensembles is beyond the scope of the present work and will be addressed separately. For the 2D model and the ion pair, where unbiased transition paths are available, the comparisons are in SI.3 and SI.4.